\documentclass[showpacs,amsmath,amssymb,preprint,superscriptaddress]{revtex4}
\usepackage{graphicx}
\newcommand{\AM}{$A\,$MeV}

\begin{document}

\title
{Isospin diffusion in semi-peripheral $^{58}Ni$+$^{197}Au$ collisions 
at intermediate energies (I): Experimental results}

\author{E.~Galichet}
\email[Corresponding author: ]{galichet@ipno.in2p3.fr}
\affiliation{Institut de Physique Nucl\'eaire, Universit\'e
Paris-Sud 11, CNRS/IN2P3, F-91406 Orsay Cedex, France}
\affiliation{Conservatoire National des Arts et M\'etiers , F-75141
Paris Cedex 03, France}
\author{M.~F.~Rivet}
\author{B.~Borderie}
\affiliation{Institut de Physique Nucl\'eaire, Universit\'e
Paris-Sud 11, CNRS/IN2P3, F-91406 Orsay Cedex, France}
\author{R.~Bougault}
\affiliation{LPC Caen, ENSICAEN, Universit\'e de Caen, CNRS/IN2P3, F-14050
Caen Cedex, France}
\author{A.~Chbihi}
\affiliation{GANIL, CEA et IN2P3-CNRS, B.P.~5027, F-14076 Caen Cedex,
  France}
\author{R. Dayras}
\affiliation{IRFU/SPhN, CEA/Saclay, F-91191 Gif-sur-Yvette, France}
\author{D.~Durand}
\affiliation{LPC Caen, ENSICAEN, Universit\'e de Caen, CNRS/IN2P3, F-14050
Caen Cedex, France}
\author{J.D.~Frankland},
\affiliation{GANIL, CEA et IN2P3-CNRS, B.P.~5027, F-14076 Caen Cedex,
  France}
\author{D.C.R. Guinet}
\author{P. Lautesse}
\affiliation{Institut de Physique Nucl\'eaire, Universit\'e Claude Bernard
Lyon 1, CNRS/IN2P3, F-69622 Villeurbanne Cedex, France}
\author{N.~Le~Neindre}
\affiliation{Institut de Physique Nucl\'eaire, Universit\'e
Paris-Sud 11, CNRS/IN2P3, F-91406 Orsay Cedex, France}
\author{O.~Lopez}
\author{L.~Manduci}
\affiliation{LPC Caen, ENSICAEN, Universit\'e de Caen, CNRS/IN2P3, F-14050
Caen Cedex, France}
\author{M.~P\^arlog}
\affiliation{National Institute for Physics and Nuclear Engineering,
RO-76900 Bucharest-M\u{a}gurele, Romania}
\author{E.~Rosato}
\affiliation{Dipartimento di Scienze Fisiche e Sezione INFN, Universit\`a
di Napoli ``Federico II'', I-80126 Napoli, Italy}
\author{B.~Tamain}
\author{E.~Vient}
\affiliation{LPC Caen, ENSICAEN, Universit\'e de Caen, CNRS/IN2P3, F-14050
Caen Cedex, France}
\author{C.~Volant} 
\affiliation{IRFU/SPhN, CEA/Saclay, F-91191 Gif-sur-Yvette, France}
\author{J.P.~Wieleczko}
\affiliation{GANIL, CEA et IN2P3-CNRS, B.P.~5027, F-14076 Caen Cedex,
  France}
\collaboration{INDRA Collaboration}
\noaffiliation

\date{\today}

\begin{abstract}

Isospin diffusion in semi-peripheral collisions is probed as a function of the
dissipated energy by studying two systems $^{58}Ni$+$^{58}Ni$ and
$^{58}Ni$+$^{197}Au$, over the incident energy 
range 52-74\AM. A close examination of the multiplicities of light
products in the forward part of phase space clearly shows an influence of the
isospin of the target on the neutron richness of these products. A progressive 
isospin diffusion is observed when collisions become more central, 
in connection with the interaction time.

\end{abstract}



\maketitle

\section{Introduction}
The knowledge of the different time scales associated to the various 
degrees of freedom involved in heavy-ion collisions at intermediate 
energy is of crucial importance to determine the physical properties 
of nuclear sources produced (in the exit channel). Thermal
equilibrium has been studied both theoretically and experimentally and 
times in the range 30-100~fm/c were derived in the Fermi energy
region~\cite{Ber78,Toe82,Cas87,Gre87,Ros89,Bor97}. With the announced 
exotic beams the N/Z degree of freedom will hopefully be explored over 
a wide range, and thus an estimate of the chemical (isospin) equilibration 
time becomes essential; moreover experimental constraints can be placed 
on the asymmetry term of the equation of state which describes its 
sensitivity to the difference between proton and neutron 
densities~\cite{Bao01,Bar02,Tsa04}. Theoretical simulations of 
collisions were performed using isospin-dependent Boltzmann-Uehling-Uhlenbeck
transport equations~\cite{Tsa04,Bao98,Shi03}. In the energy domain 20-100\AM, 
estimates of the chemical equilibration times in the range 40-100~fm/c are 
reported, if one excludes calculations with an asymmetry term rapidly
increasing around normal density.

Experimentally some investigations concerning this time scale have been done.
 In the Fermi energy domain, studies on isospin equilibration in fusion-like
 reactions between medium nuclei (A$\sim50$) have shown that isospin
 equilibrium occurred prior to light fragment emission, which gives
 an upper limit  around 100~fm/c~\cite{johnston1,johnston2};  for peripheral 
 collisions between Sn  isotopes~\cite{Tsa04}  only a partial
 equilibrium (isospin asymmetry of the projectile remnant is
half way between that of projectile and the equilibration value) is measured
at the separation time ($\sim$100~fm/c) between quasi-projectile and
quasi-target. At higher incident energy (400\AM) 
the FOPI Collaboration measured the degree of isospin mixing between projectile
and target nucleons, and found that complete mixing is not 
reached even in the most central collisions~\cite{rami}.

In the present study we concentrate on semi-peripheral collision measurements
performed with the INDRA array. 
The properties of the de-excitation products of the quasi-projectiles
inform on the degree of N/Z diffusion and the separation time
between the two partners will be taken 
as a clock to derive qualitative information on isospin equilibration. Two reactions with the same 
projectile, $^{58}Ni$, and two different targets ($^{58}Ni$ and $^{197}Au$) 
are used at incident energies of 52\AM{} and 74\AM.
The N/Z ratios of the two systems are 1.07 for Ni+Ni 
and 1.38 for Ni+Au. INDRA only provides isotopic identification up to 
beryllium and does not detect neutrons. Thus an N/Z ratio for complex 
particles is constructed which well reflects the evolution of the N/Z of quasi-projectiles with the 
violence of the collisions. (see the accompanying article~\cite{galichet2}).
The Ni+Ni symmetric system is taken as a 
reference since, on average, the isospin should remain
constant with time whatever the collision process is.

The paper is divided into three sections.
In a first part, we describe the experiment and the event selection,
then we present the properties of quasi-projectiles and finally
we discuss the evolution of the isospin, before concluding.

\section{Experiment and event selection}\label{expsel}

\subsection{Experimental details}\label{exp}

 $^{58}$Ni projectiles accelerated to 52 and 74\AM{} by the GANIL facility
impinged on $^{58}$Ni (179 $\mu$g/cm$^2$) and $^{197}$Au (200 $\mu$g/cm$^2$)
targets. The charged products emitted in collisions were collected
by the $4\pi$ detection array INDRA. A detailed description of the apparatus
can be found in references~\cite{pouthas1,pouthas2,steck}. All elements were
identified within one charge unit up to the projectile charge. Elements
from H to Be were isotopically separated when their energy was high enough
(above 3, 6, 8~MeV for p, d, t; 20-25~MeV for He isotopes; $\sim$60~MeV for Li
and $\sim$80~MeV for Be).
However isotopic identification was not possible in the first ring of INDRA,
constituted of phoswiches, so in this paper the angular range is limited to
3-176$^o$ for all products.
In the following we shall call fragments the products for which only the
atomic number is measured (Z$\geq$5).
The on-line trigger required that four modules of the array fired. 
The off-line analysis only considered
events in which four charged products were identified.

\begin{table}[hbt]
\caption{\label{sys} (color online) Characteristics of the systems studied : grazing angle,
reaction cross section (calculated from~\cite{Kox84}), and measured cross
sections after the different selections}
\centerline{\begin{tabular}{|l|c|c||c|c|}
\hline
 & \multicolumn{2}{c||}{Ni + Ni} & \multicolumn{2}{c|}{Ni + Au} \\
E$_{inc}/A$ (MeV)& 52 & 74 & 52 & 74 \\
E$_{c.m.}$ (MeV) & 1508 & 2146 & 2330 & 3316 \\
\hline
$\theta_{gr}$ (lab)& 1.9$^o$ & 1.3$^o$ & 4.6$^o$ & 3.2$^o$  \\
\hline
$\sigma_R$ (mb)         & 3460 & 3410 & 5400 & 5400 \\
$\sigma_{M\geq 4}$ (mb) & 1553 & 1634 & 3780 & 3807\\
Selected events (mb)    & 1032 &  953 & 3034 & 2885 \\
Selected QP (mb)        &  624 &  491 & 904  &  793 \\
\hline   
\end{tabular}}
\end{table}
The characteristics of the systems studied here are displayed in
table~\ref{sys}. Note that the grazing angle is below the minimum detection
angle of INDRA (2$^o$) for the Ni+Ni system at both energies. 
This shows through the lower
measured percentage of the reaction cross sections ($\sigma_{M\geq 4}$,
table~\ref{sys}) for the Ni+Ni system (around 50\%) as compared to those for 
the Ni+Au system (70\%).
The measured cross sections are derived from target thicknesses and
integrated beam fluxes. 

\subsection{Event selection}\label{selevt}
A first and simple selection required that the total detected charge 
amounts to at least 90\% of the charge of the projectile. 
\begin{figure}[htbp]
\resizebox{0.6\textwidth}{!}{%
\includegraphics{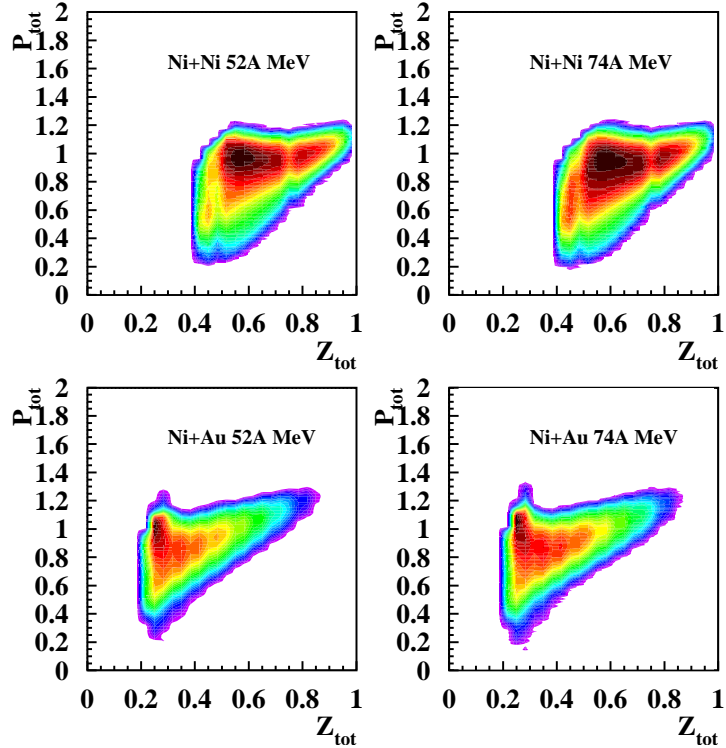}}
\caption{(color online) The detected momentum $P_{tot}$ versus the
total charge $Z_{tot}$ normalized to the incident momentum and the total 
charge of the system, for the Ni+Au and Ni+Ni systems at 52 and 74\AM{} for
the first selection. Event scale is logarithmic.}
\label{figure1}
\end{figure}
 Figure~\ref{figure1} 
shows, for the two systems and the two energies, the location of the 
selected events in the total detected charge and momentum plane. 
In table~\ref{sys}, one can observe that after this event selection, 
about 30\% of the reaction cross section is kept for Ni+Ni, against 55\% 
for the Ni+Au system: because of the detection geometry, 
peripheral collisions (Z$_{tot}$ $\sim$ Z$_{proj}$ and P$_{tot} \sim$1)
are drastically suppressed in the Ni+Ni reactions, 
neither the projectile nor the target remnants are detected. 
This effect exists but to a lesser extent for the very asymmetric Ni+Au
system, thanks to the 
larger value of the grazing angle: here events with 
Z$_{tot}$ $\sim$ Z$_{proj}$ and P$_{tot} \sim$1 are clearly visible.
\begin{figure}[htbp]
\resizebox{0.6\textwidth}{!}{%
\includegraphics{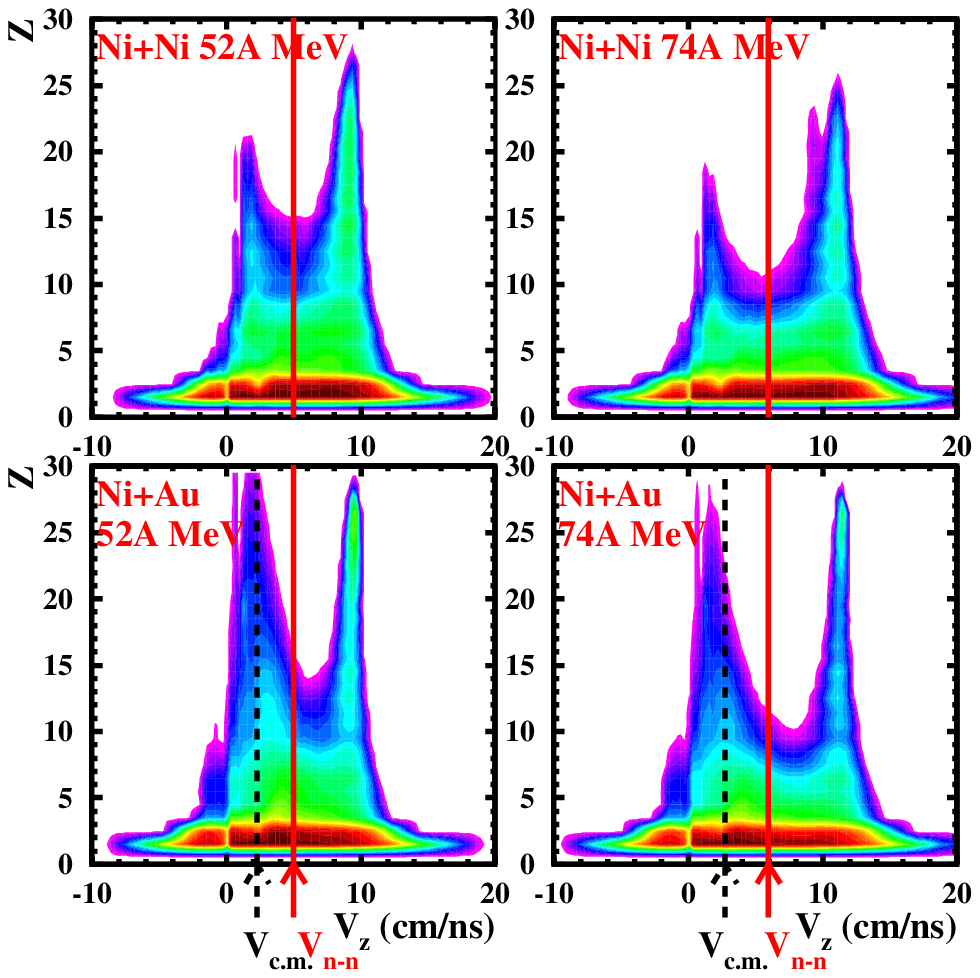}}
 \caption{(color online) The emitted products, for the selected events,
in $Z - V_{Z}$ plane for the Ni+Ni and Ni+Au sytems at 52 and 74\AM. V$_Z$
is in the laboratory system.}
\label{figure2}
\end{figure}
Conversely for this system the probability to
detect all the products of an event (Z$_{tot}$ $>$ 0.6) is very small: 
the target-like fragment remains generally undetected because of the 
thresholds, unless it undergoes fission. 

\subsection{Selection of the quasi-projectile}\label{selqp}

A further selection must be done to select the ``quasi-projectile''.
 We do not intend to isolate a ``source'', but rather to select a
forward region in phase space where the detected products have a small
probability to result from emission by the quasi-target. In principle, 
this could be done
by a cut at the center-of-mass velocity; for the asymmetric system however, 
the target being more than three times heavier than the projectile, some
particles from the target would be kept, as seen in fig.~\ref{figure2}
which shows the charge of the products as a function of their laboratory 
velocity along the beam axis for the selected sets of events.
Thus the cut was made at the nucleon-nucleon velocity - note that both 
cuts are identical for the Ni+Ni system.
The quasi-projectile selection only
keeps particles and fragments with a parallel velocity higher than
the nucleon-nucleon velocity. It was verified that in this region of
velocity space, all isotopes of H up to Be were fully identified (Z and A).
In fig.~\ref{figure2} a small contribution of fragments which have a 
velocity $\sim$10\% smaller 
than the projectile velocity appears for the Ni+Ni reaction at 74\AM{}
incident energy. It was attributed to a beam halo interacting with the 
brass target holder and represents about 14\% of the events~\cite{guy}.
In order to sharpen the comparison between quasi-projectiles produced 
in the two systems, the total charge beyond the nucleon-nucleon velocity 
was required to be in the range 24-32.
In all cases 1.3 to 2$\times 10^6$ events are kept,
amounting to 14-18\% of the reaction cross sections.

\begin{figure}[htbp]
\resizebox{0.6\textwidth}{!}{%
\includegraphics{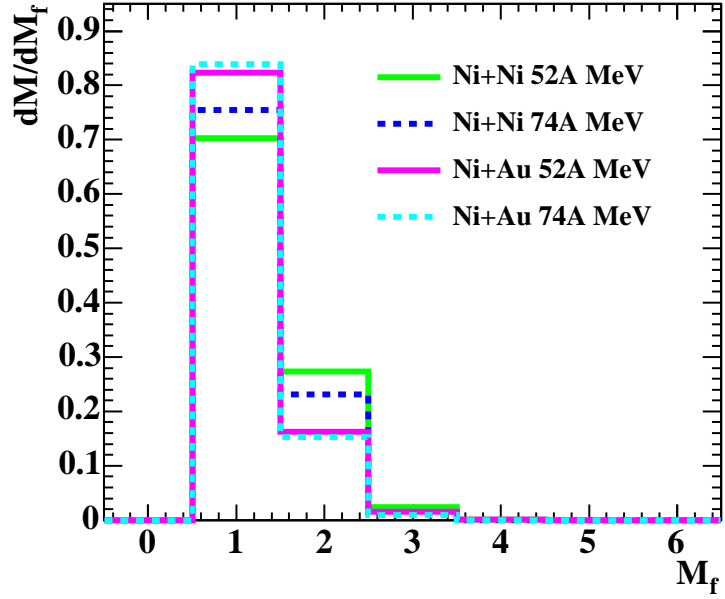}}
 \caption{(color online) Multiplicity distribution for
fragments, $Z\geq5$, for the selected quasi-projectiles.}
\label{figure3}
\end{figure}

In short in the following we call ``quasi-projectile'' the ensemble of 
charged products 
which have a velocity higher than the nucleon-nucleon velocity, without
prejudice on the shape, degree of equilibration {\ldots} of the ensemble so
defined.
In  figure~\ref{figure3} is represented the
 fragment ($Z \geq 5$) multiplicity distribution after all selections.
In all cases a majority of the quasi-projectiles have only one fragment, which
can be considered as the quasi-projectile remnant. For the Ni+Ni system,
about 25-30\% of the events have two or more fragments 
while for the Ni+Au system this percentage is smaller ($\sim$15\%).

\section{Properties of the quasi-projectiles}\label{propqp}

\subsection{Event sorting}\label{excit}

\begin{figure}[htbp]
\resizebox{0.6\textwidth}{!}{%
\includegraphics{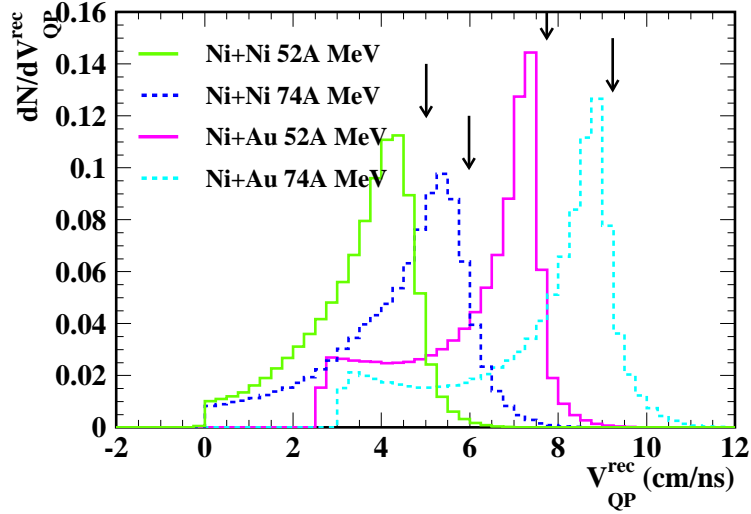}}
\caption{(color online) Reconstructed velocity of the quasi-projectile 
in the center of mass frame for the two systems and the two energies.
The arrows indicate the projectile velocity.}
\label{figure4}
\end{figure} 

The method consisting in doing a calorimetry from
the measured products can not be applied here, because it would require
firstly to isolate sources and secondly an
assumption on the number of neutrons, and thus on the N/Z of the
quasi-projectiles, which is just the quantity we want to work out.
To avoid the above difficulties we choose to sort the events as a function
 of the dissipated energy, calculated in
 a binary hypothesis, with the assumptions detailed
below. \\
i) The quasi-projectile velocity is equal to
the measured velocity of the fragment, or reconstructed from the velocity of
all the fragments it contains.
The distribution of the quasi-projectile velocities ($V_{QP}^{rec}$) 
so determined are represented, in the center-of-mass
reference frame,  on figure~\ref{figure4}.
In all cases the reconstructed velocity of the quasi-projectile peaks at a
value smaller than the projectile velocity, but remains closer to it for 
the Ni+Au system.  
\begin{figure}[htbp]
\resizebox{0.6\textwidth}{!}{%
\includegraphics{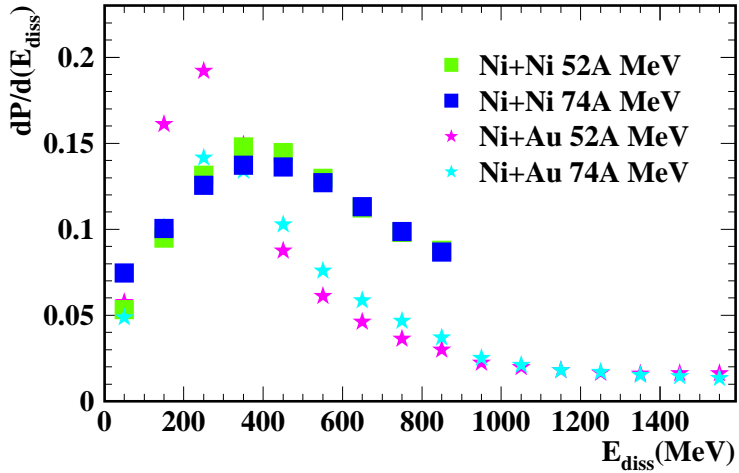}}
\caption{(color online) Distributions of the dissipated
energy for Ni+Ni and Ni+Au at 52 and 74 \AM.}
\label{figure5}
\end{figure}

ii) The relative velocity between the quasi-projectile and the quasi-target
is determined as if the collision was purely binary, without mass exchange:
\begin{equation} \label{eq:vrel}
V_{rel}=V_{QP}^{rec} \times \frac{A_{tot}}{A_{target}}
\end{equation} 
and thus the total dissipated energy reads:
\begin{equation} \label{eq:Eexc}
E_{diss}=E_{c.m.}-\frac{1}{2}\mu V_{rel}^2 ,
\end{equation}
with $\mu$ the initial reduced mass.
It is demonstrated in~\cite{Yan03,Pian} that the velocity of the QP is a good 
parameter for following the dissipated energy, except in very peripheral
collisions, due to trigger conditions. Moreover, it is shown in figure 5 of
the accompanying paper that $E_{diss}$ gives a good measure of the impact
parameter.

In figure~\ref{figure5} are represented the dissipated energy distributions. 
For the Ni+Ni system and at the two incident energies,
the distributions present a maximum at 
E$_{diss}\approx$ 350 MeV, while they peak at lower dissipated energies for
Ni+Au. This was expected from the remarks made in the previous sections; 
the most peripheral collisions - low excitation energies - are much more 
poorly sampled  for Ni+Ni reactions than for Ni+Au. The comparison of the 
properties of the quasi-projectiles between the two systems will be  made 
by sorting data in  bins of 100~MeV dissipated energy.

\begin{figure}[htbp]
\resizebox{0.7\textwidth}{!}{%
\includegraphics{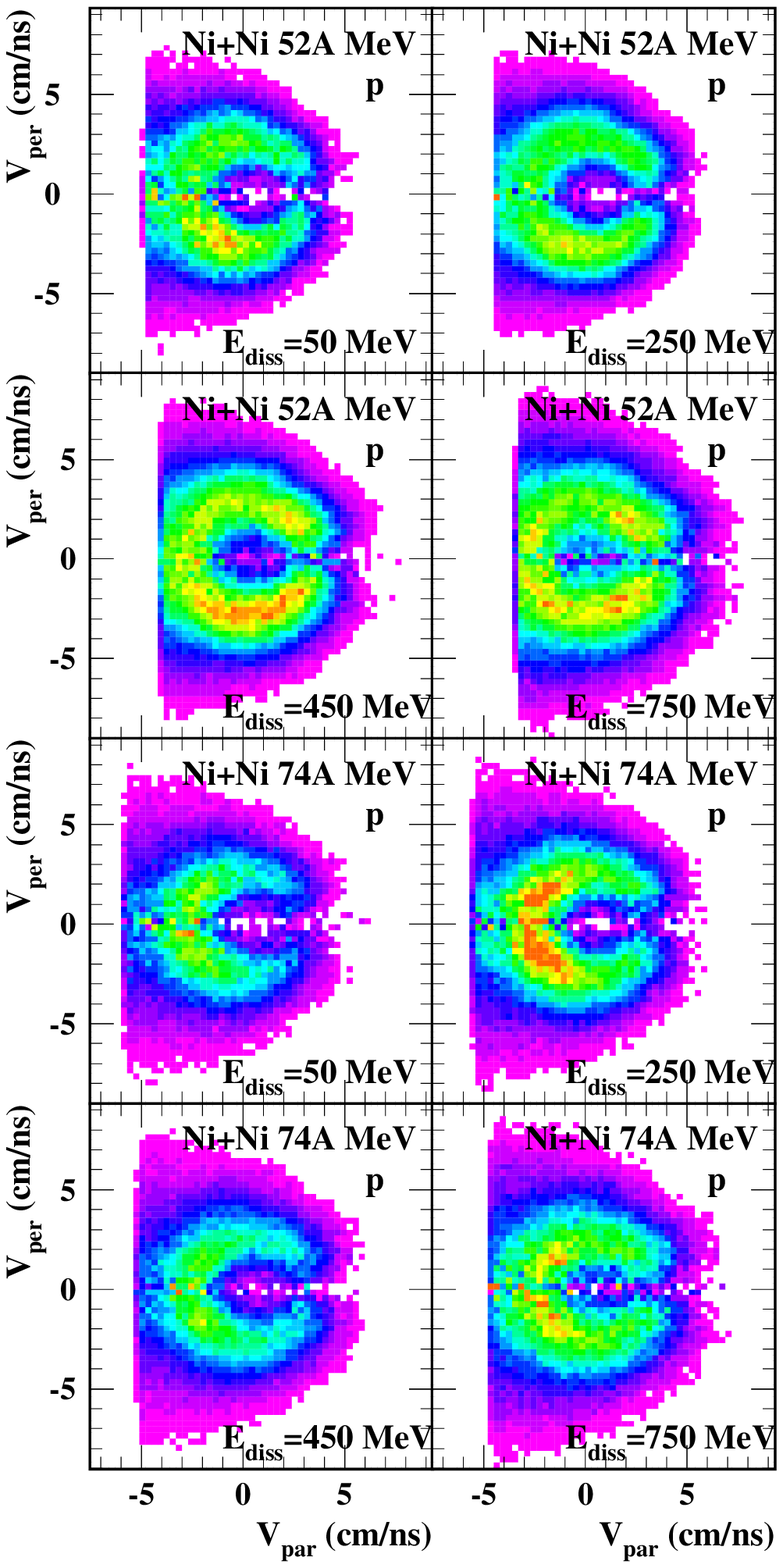}
\includegraphics{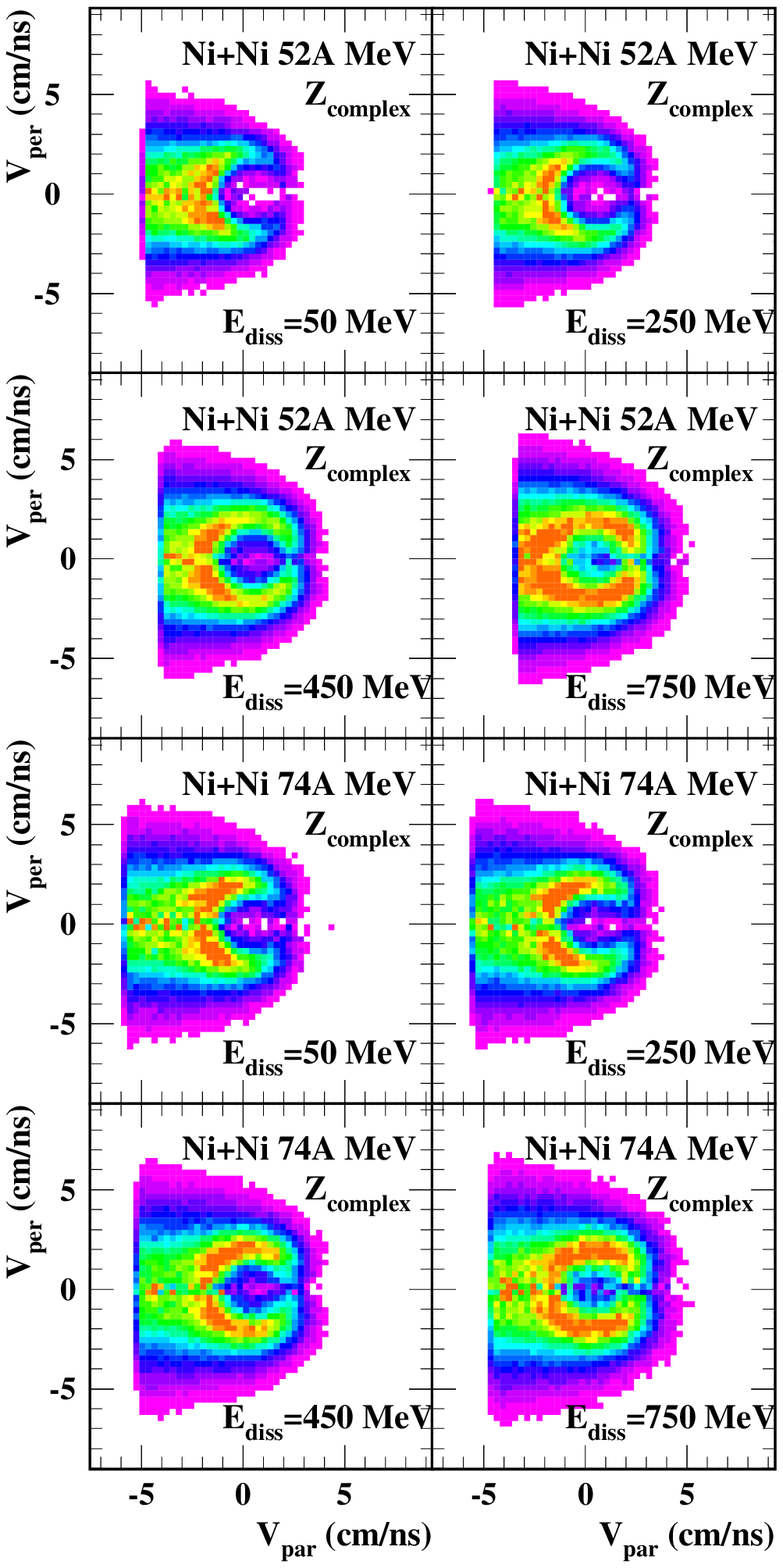}}
\caption{(color online) Invariant cross sections for 
protons (left) and complex particles (right) emitted in the Ni+Ni system 
at 52 and 74\AM. Velocities are 
expressed in the quasi-projectile frame. Contour levels are equidistant.}
\label{figure6}
\end{figure} 

\begin{figure}[htbp]
\resizebox{0.7\textwidth}{!}{%
\includegraphics{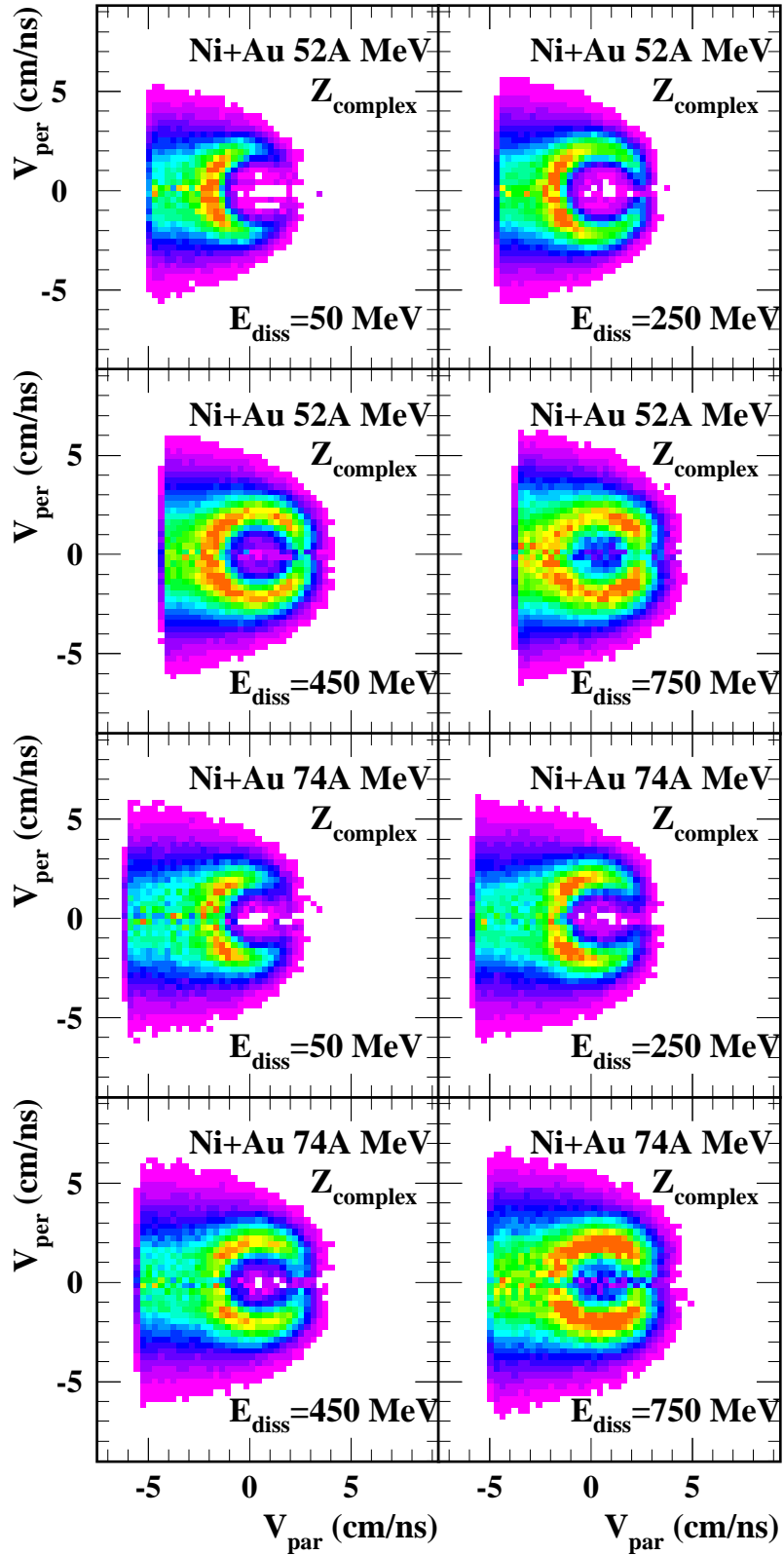}}
\caption{(color online) Invariant cross sections of complex particles
emitted in Ni on Au collisions at 52 and 74\AM.}
\label{figure8}
\end{figure} 

\subsection{Invariant cross-section plots}
As a verification of the selections and sorting made, we examined the
repartition of the different particles in the velocity plane.
A sign has been attributed to the perpendicular velocity
depending on the value of the azimutal angle ($V_{per}<0$ corresponds to
azimutal angles larger than 180$^o$). Such plots in the lab system allow
a rough verification of the good operation of the INDRA array. 
Figs.~\ref{figure6},~\ref{figure8} are presented in the QP frame. 
The observed asymmetry between positive and negative values of $V_{per}$
comes from a deviation of the beam position from the symmetry axis of INDRA,
reflected in the azimutal distribution of projectile residues, and thus 
showing up when transforming all particles in the frame of this fragment. 
This is particularly visible for protons at 52\AM{} in fig.~\ref{figure6}, 
and does not affect the following results.
In the left panel of figure~\ref{figure6} the invariant cross sections 
for protons emitted in the Ni+Ni reactions are presented.
For all bins of dissipated energy and for the two incident energies, 
well-defined Coulomb circles are visible, showing that the protons
essentially come from one source. 
The mid-rapidity/neck emission does not seem to be prominent at 52\AM, 
except at low dissipation, 
(due to the online trigger, when QP and QT are too little excited 
for evaporating charged particles, configurations with several mid-rapidity
particles are enhanced)
while it becomes more important for all dissipations 
at 74\AM. Due to the smaller quasi-projectile velocity at 52\AM, and to 
the large proton velocities, the Coulomb circles are slightly cut at the 
higher dissipated energies (upper pannel). On the right panel of 
figure~\ref{figure6} are displayed the same plots for complex particles,
including deuterons, tritons, helium, lithium and beryllium isotopes
and labelled $Z_{complex}$ in figures~\ref{figure6} and~\ref{figure8}.
The Coulomb circles are also clearly visible, but an accumulation of 
particles appears  backwards of the quasi-projectile due to the
importance of mid-rapidity emission for such products; for the highest
dissipated energies the distributions become more forward/backward symmetric
for particles emitted at the Coulomb velocity.
No sizeable emission from the target is present. 

For protons emitted in Ni+Au collisions, the pictures (not shown) resemble
closely those for Ni+Ni at the same incident energy.
In figure~\ref{figure8}  are represented the same
pictures for complex particles for the Ni+Au system. 
As in the Ni+Ni system neck emission is apparent at low dissipation
and Coulomb rings become more symmetric for higher dissipated
energies in all cases.

\subsection{The heaviest fragment}
\begin{figure}[htbp]
\resizebox{0.6\textwidth}{!}{%
\includegraphics{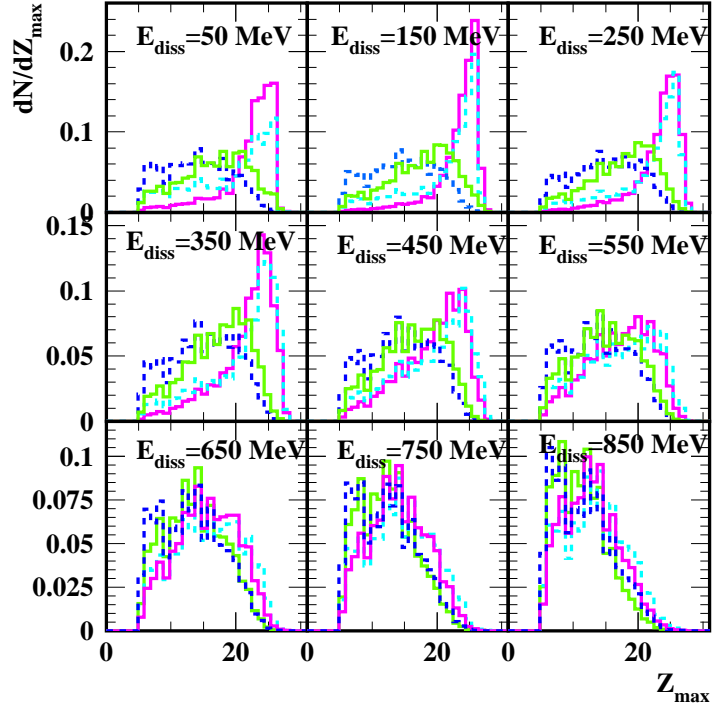}}
\caption{(color online) Distribution of the heaviest 
quasi-projectile fragment for the four systems.
The line codes are the same as in fig.~\ref{figure4}}
\label{figure9}
\end{figure}
Figure~\ref{figure9} represents the charge distribution of the
heaviest fragment  ($Z\geq 5$) of the quasi-projectile, for the four
reactions in each dissipated energy bin. For the Ni+Au system, the heaviest
quasi-projectile fragment has a charge around Z$_{max}$=24 for peripheral 
collisions (E$_{diss} <$ 450~MeV).  These events have a quasi-projectile 
fragment multiplicity of one. In all other cases for this system there is 
no privileged value of the maximum charge, they often correspond to
quasi-projectiles with more than one fragment.  Note that  the 
Z$_{max}$ distributions barely depend on the target or on the energy when
the dissipated energy overcomes 600~MeV.

For the Ni+Ni system, the distributions of Z$_{max}$ do not exhibit any
peak whatever the dissipation, which again agrees with the lack of very 
peripheral collisions in the event samples.

\subsection{Multiplicity of particles}

\begin{figure}[htbp]
\resizebox{0.7\textwidth}{!}{%
\includegraphics{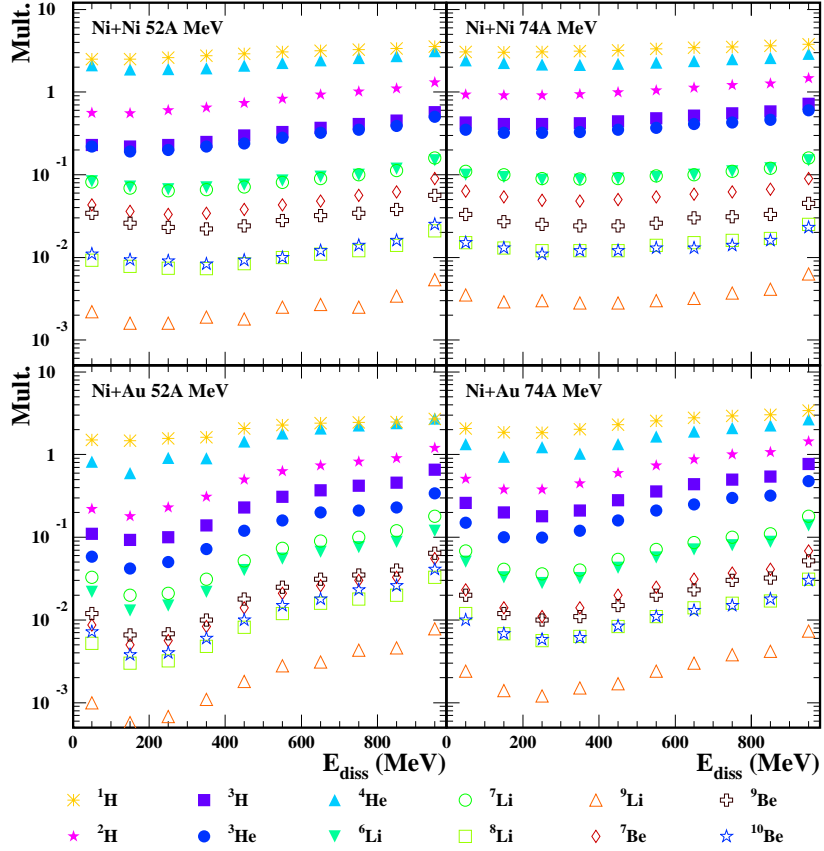}}
 \caption{(color online) Multiplicity of emitted products
($Z<5$) versus the dissipated energy. In all cases, statistical error bars
are smaller than the size of the symbols}
\label{figure10}
\end{figure}
The average multiplicities of isotopically resolved charged products 
associated with quasi-projectiles for each  energy bin are displayed in 
figure~\ref{figure10}.

Let us first examine the Ni+Ni system. For hydrogen isotopes the
multiplicity is constant at low energies and then rises. For
other products the multiplicity slightly decreases before rising with the
dissipated energy, above 400-500 MeV. It is indeed at this value that the 
dissipated energy distribution of the selected quasi-projectiles shows
a maximum (fig~\ref{figure5}). The effect is more marked for the
heavier products; indeed 
for the most peripheral collisions, due to the on-line trigger (4 detected
charged products), some particular configurations were retained namely those
with the highest multiplicities. A  situation sampling all
configurations for a given dissipation is recovered when the multiplicities
start to increase.
At 52\AM{} the ratio between the maximum and the minimum values of the
multiplicities is around 1.5 for protons and $\alpha$ particles, and more
than 2.5 for other products. At 74\AM, the observed multiplicities are
generally close to those for 52\AM. Systematically higher values are
however found at lower dissipated energies ($<$400 MeV) and in all cases for
neutron rich hydrogen and lithium. This obviously comes from the sorting
parameter which is not the excitation energy of an equilibrated piece of
matter. At the higher energy the ratios between the maximum and the minimum
values of the multiplicity of any species are smaller than at 52\AM.

For the Ni+Au system  
the multiplicity variations closely resemble those of the Ni+Ni case, showing 
firstly a slight decrease before a neat increase above dissipations of
250~MeV, which also corresponds to the peak in the dissipated energy 
distribution (see fig.~\ref{figure5}).
At both incident energies the ratios between the maximum and the minimum 
values of the multiplicity of any species are much higher than in the 
corresponding Ni+Ni case, particularly for lithium and beryllium isotopes 
(ratios as high as 10 are observed).  The multiplicities for neutron rich
species are smaller at 74 than at 52\AM, showing the reverse evolution
with respect with the Ni+Ni data.
If one now compares the different multiplicities for Ni+Ni and Ni+Au, several
differences immediately appear from fig~\ref{figure10}: for Ni+Au, there are
twice more protons than $\alpha$'s at low dissipated energy, 
while both multiplicities tend towards equal values with increasing
dissipation. Conversely the difference between
proton and $\alpha$ multiplicities is almost constant  around 30 (40)\%
for Ni+Ni at 52 (74)\AM. For the Ni+Au system, all neutron rich isotopes
are more abundantly produced, as can be seen by comparing tritons
and $^3$He, $^6$Li and $^7$Li, $^7$Li and $^7$Be, $^7$Be and $^9$Be.
A simple way of observing the isospin effect is to calculate the average mass
per element for the two systems, starting from figure~\ref{figure10}.
As expected, and due to the huge dominance
of $\alpha$'s, the average mass of helium is insensitive to the isospin of
the target at variance to those of hydrogen, lithium and beryllium which
increase with the target neutron excess. These observations indicate that there
is a transfer of neutrons, or an isospin diffusion, from the backward to the
forward part of phase space. Similar observations were made for vaporised
silicon quasi-projectiles after interaction with targets with different
isospins~\cite{Ves00}; as in this paper, we also notice, from the evolution
of the average element masses, that hydrogen and beryllium are more
sensitive to the isospin of the target than lithium.
The average masses are however insensitive to the dissipation, except a
slight increase observed for hydrogen in Ni+Au data. A combination of the 
multiplicities of the different light isotopes will therefore bring more
information than the individual evolution per element.
The authors of ~\cite{VesR00} also noted a
decrease of the t/$^3$He ratio at low temperature, as predicted by Lattice
Gas Model calculations~\cite{Cho99}. In the present data the evolution of
this ratio with excitation is weak but follows the same trend.

To summarize this part, a close examination of the multiplicities of light
products in the forward part of phase space clearly shows an influence of the
isospin of the target on the neutron richness of these products. In other
words there is an isospin diffusion from the target side to the projectile
side in the course of the reaction. This effect will be quantified by a
single variable in the next section.

\section{Isospin diffusion and equilibration} 

\subsection{Isospin ratio of complex particles}

The isospin ratio of quasi-projectiles in intermediate energy heavy ion
collision was abundantly studied in the 80's, when the first beams at these
energies appeared. The underlying idea was already the determination of the
equilibration time of the isospin degree of freedom. The reactions involved
 $^{40}$Ar and $^{84,86}$Kr projectiles at 27-30 and 44\AM{}
(see~\cite{Bor90} and references therein for a review). The average N/Z were
determined from Z=5 up to the projectile charge, at very forward angles,
thus for very small dissipation; to our knowledge, no attempt to study the
evolution of N/Z as a function of the dissipated energy, as proposed in the
present paper, was ever made.
 For a given projectile and bombarding energy, it was found that
the average N/Z of
residues increases with the target N/Z. The difference between a $^{58}$Ni
and a Au target becomes smaller when the incident energy increases. Indeed
the average N/Z tends towards that of the valley of stability, because of
increasing dominance of the de-excitation process. 
This indicates that to characterize the
primary process, not only the projectile residue (Z$_{max}$ in this paper),
but also all the emitted products should be detected, including neutrons.
No data ever reached this ultimate goal. More information should however be
extracted from the emitted products than from Z$_{max}$. 
In the experiments discussed here,
INDRA only provides isotopic identification for
isotopes of hydrogen up to beryllium and moreover does not detect neutrons.
We wanted to avoid any hypothesis on heavy fragment masses and on the
number of emitted neutrons, which would bias our conclusions; we thus
construct an isospin ratio for complex particles, most probably different
from the (N/Z) of the quasi-projectile, but evolving in the same way with
increasing dissipation,
as it is shown in the joint paper~\cite{galichet2}.
This variable, ($<N>$/$<Z>$)$_{CP}$, is calculated for each dissipated 
energy bin (containing $N_{evts}$ events) and is defined as

\begin{equation}
(<N>/<Z>)_{CP} = \sum_{N_{evts}}{\sum_{\nu} {N_{\nu}}}
/ \sum_{N_{evts}}{\sum_{\nu} {P_{\nu}}}
\end{equation}

where $N_{\nu}$ and $P_{\nu}$ are respectively the numbers of neutrons and
protons bound in particle $\nu$ , $\nu$ being d, t, $^3$He, $^4$He, $^6$He,
$^6$Li, $^7$Li, $^8$Li, $^9$Li, $^7$Be, $^9$Be, $^{10}$Be; free protons are
excluded as well as $^8$Be, the latter because they are only partly 
identified, when the two $\alpha$'s that they emit hit the same
scintillator.
The relative abundances of these nuclei among all those emitted by
the quasi-projectiles are assumed to reflect the isospin of the initial
emitter. 
We recall that the light nuclei included in eq.~3 are fully
identified, without any energy threshold. 
Relative systematic errors on ($<N>$/$<Z>$)$_{CP}$
as a function of dissipation mainly come from the wrong identification of a
$^8$Be as two $\alpha$'s; they are lower than 0.4\%.

\subsection{Evolution of isospin with centrality}

\begin{figure}[htbp]
\resizebox{0.6\textwidth}{!}{%
\includegraphics{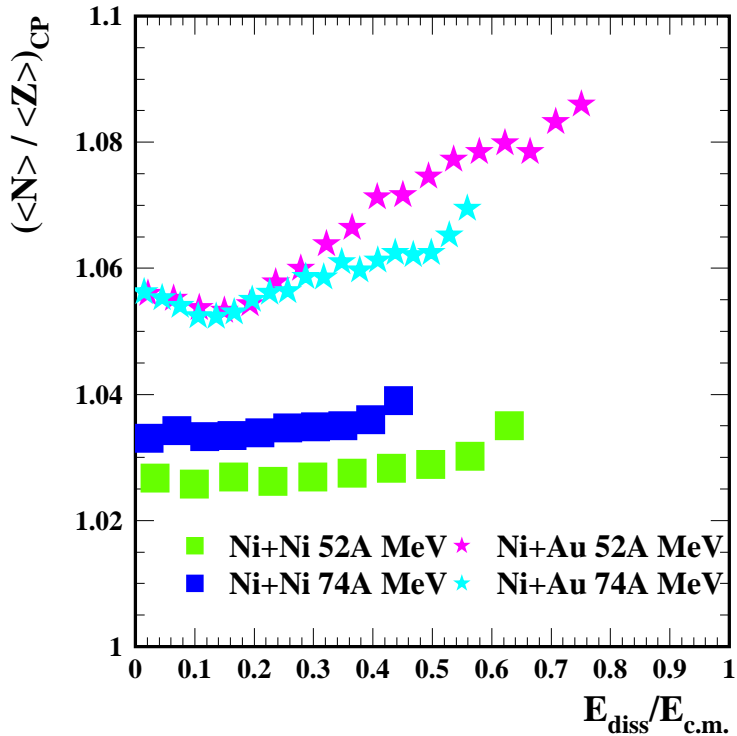}}
\caption{(color online) Isospin ratio for complex particles as a function 
of the normalised dissipated energy. 
Statistical error bars are smaller than the size of the symbols.}
\label{figure11}
\end{figure} 
In figure~\ref{figure11}  the N/Z ratio for complex particles is plotted 
as a function of the normalised dissipated energy for the four reactions.
The stars correspond to the Ni+Au system and the squares to the Ni+Ni 
system. 
We stress again that errors are very small,
and thus the observed variations are significant.
It immediately appears from the figures  that 
the behaviour of ($<N>$/$<Z>$)$_{CP}$ is completely different for the two 
systems.

Let us first examine the Ni+Ni data.  Within about 1.5\% ($<N>$/$<Z>$)$_{CP}$
values are independent of the dissipated energy, at 52 and 74\AM;
this can be interpreted as the sign that the variable, used as reference,
well reflects the 
evolution of the 
average isospin of the initial quasi-projectiles,
which is expected to be constant for this system, provided that
the de-excitation
process does not influence the isospin ratio. The observed value of
($<N>$/$<Z>$)$_{CP}$,  closer to 1 than the N/Z of the system (1.07),
comes from 
the dominance of $\alpha$'s  among the particles used to calculate it.
Moreover ($<N>$/$<Z>$)$_{CP}$ is little dependent on
the incident energy. 
The slight difference observed must be attributed to direct particles 
emitted at mid-rapidity or neck effect, which are included in the calculation 
of ($<N>$/$<Z>$)$_{CP}$. Another explanation may be that
the system is proton rich, which favors proton preequilibrium
emission; preequilibrium emission is expected to increase with the 
incident energy 
leading to an increase of the N/Z of primary
quasi-projectile (target). This effect is observed in  the isospin ratio,
independently of the dissipation as the system is expected to remain, on
average, symmetric. Observing such an effect is an indication that the 
chosen variable is indeed sensitive to the initial N/Z of the quasi-projectile.

Let us turn now to Ni+Au.
A first observation is that the isospin ratio of the Ni+Au system is higher
than the isospin ratio of the Ni+Ni system whatever the dissipated energy.
One may argue that the difference of ($<N>$/$<Z>$)$_{CP}$
between the two systems is small (0.02-0.05, compared to 0.31 for the true
N/Z values for the composite systems). This again can be attributed to the
definition of the variable, built on particles among which deuterons
and $\alpha$'s are dominant. Therefore
($<N>$/$<Z>$)$_{CP}$ remains closer to 1 than the true isospin of the
quasi-projectiles, the larger excess of neutrons being evacuated by
free neutrons.
The heavy fragments do not carry away a lot of neutrons :
it was shown in~\cite{Day86} that for a given projectile, the $<N>$/$Z$
of the heavy fragments are more neutron rich with a Au target than with a Ni
target: the difference in average N/Z for $^{40}$Ar residues is about
0.02-0.03, namely the same as what is observed here.
Owing to these effects, we think that the gap between the values of
($<N>$/$<Z>$)$_{CP}$ for the two systems is significant and the mixing with
the Au target did occur.
Hence our variable well reflects the isospin diffusion
between the target and the projectile
as it is confirmed by figure 8 of the
joint paper~\cite{galichet2} where one can see that ($<N>$/$<Z>$)$_{CP}$ values change as
those of N/Z for quasi-projectiles but within a reduced and lower domain.
For the three first bins of 
dissipated energy  ($<N>$/$<Z>$)$_{CP}$ has the same value for the two incident
energies and slightly decrease with dissipation.
These points should however be regarded with caution, they correspond to the
region where multiplicities decrease with increasing dissipation, due to 
the selection of particular configurations by the on-line trigger.
The isospin ratio of the Ni+Au system is however higher than 
that of the Ni+Ni system, which could arise from
the neutron skin of the Au target and/or from the mid-rapidity
particles included in our quasi-projectile selection which are more neutron 
rich~\cite{I23-Lef00,I17-Pla99}.
This result is a first indication of isospin diffusion.

At higher dissipated energies, ($<N>$/$<Z>$)$_{CP}$ behaves differently
depending on
the incident energy. While ($<N>$/$<Z>$)$_{CP}$ presents a significant increase
with dissipation at 52\AM, the trend is flatter at 74\AM.
($<N>$/$<Z>$)$_{CP}$ thus reaches higher values at 52\AM.
This may be interpreted as a progressive isospin diffusion when collisions
become more central, in connection with the interaction time. For a given
centrality, the separation time is longer at 52\AM{} than at 74\AM,
leaving more time to the two main partners to go towards isospin equilibration.

\section{Conclusion}

To summarize, the value of the isospin variable
($<N>$/$<Z>$)$_{CP}$ for Ni+Au is different from and larger than
that for Ni+Ni.
It does not significantly evolve for the Ni+Ni system, neither with the excitation energy,
nor with the incident energy when increased from 52 to 74\AM.
Therefore ($<N>$/$<Z>$)$_{CP}$ for Ni+Ni provides a good reference
to which the same
variable for Ni+Au can be compared. The continuous increase of
($<N>$/$<Z>$)$_{CP}$ up to
the highest observed dissipation for Ni+Au at 52\AM{} indicates that at least
a partial isospin equilibration is reached at the corresponding separation
time, $\sim$80~fm/c. The separation time $t_{sep}$ was estimated by
$t_{sep}$$\sim$$(D_{Ni}+D_{Au}+d)/v_{beam}$$\sim$80~fm/c at 52\AM{} and
66~fm/c at 74\AM; $D$ is the nuclear diameter, $v_{beam}$ the incident
velocity and $d$=3~fm the distance between the two nuclear surfaces at 
separation. \\
We will see in the accompanying article~\cite{galichet2} that, in the 
framework of the model
employed, ($<N>$/$<Z>$)$_{CP}$ gives a reliable picture of
isospin diffusion in the reactions studied and is relevant to determine if
isospin equilibration takes place or not for the high excitation energies.


\end{document}